\documentclass{ws-ijmpa}
\usepackage{epsfig}
\usepackage{axodraw}

\begin{document}

\markboth{Authors' Names}
{Instructions for Typing Manuscripts (Paper's Title)}

\catchline{}{}{}{}{}
%
%%%%%%%%%%%%%%%%%%%%%%%%%%%%%%%%%%%%%%%%%%%%%%%%%%%%%%%%%%%%%%%%%%%%

\title{``Double Chargino Production in $e^{-}e^{-}$ scattering"}

\author{M. C. Rodriguez}

\address{Funda\c c\~ao Universidade Federal do Rio Grande-FURG \\
Departamento de F\'\i sica \\
Av. It\'alia, km 8, Campus Carreiros \\
96201-900, Rio Grande, RS \\
Brazil\\$^*$E-mail: mcrodriguez@fisica.furg.br}

\maketitle

\begin{history}
\received{17 May 2007}
\revised{Day Month Year}
\end{history}

\begin{abstract}
We point out the production 
of the charginos and neutralinos in electron-electron process in several supersymmetric 
models, in order to show that the International Linear Collider can discover 
double charged charginos if these particles really exist in nature.

\end{abstract}

\ccode{PACS numbers: 12.60.-i %Models beyond the standard model
12.60.Jv %Supersymmetric Model
13.85.Lg %Total cross section
}

\section{Introduction}

The full symmetry of the so called Standard Model (SM) is the gauge group
$SU(3)_{c}\otimes SU(2)_{L}\otimes U(1)_{Y}$. This model describes the 
observed properties of charged leptons and quarks it is not the ultimate 
theory. However, the necessity to go beyond it, from the 
experimental point of view, comes at the moment only from neutrino 
data. If neutrinos are massive then new physics beyond the SM is needed.
From the theoretical point of view, the SM cannot be a fundamental theory 
since it has so many parameters and some important questions like that of 
the number of families do not have an answer in its context.

On the other side, it is not clear what the physics beyond the SM should be.
Probably, the SM is an effect of grand unified scenarios and/or their
supersymmetric extensions, the Minimal Supersymmetric Standard Model (MSSM) \cite{mssm}. 

There 
are two Higgs doublets in the MSSM, the Higgs' Mass spectrum was studied at\cite{INO82a,INO82b}. 
The Higgs sector of the MSSM is established by the charged Higgs bosons~($H^{\pm}$),
the neutral Higgs bosons $H^{0}$, $h^{0}$ and $A^{0}$ and finally the charged~($G^{\pm}$) 
and neutral Goldstone bosons~($G^{0}$). The upper limit on the
mass of the lightest neutral scalar is lighter than $M_Z$ at the tree level but 
radiative corrections rise it to 130 GeV~\cite{haber2}.

By another hand, the main motivation to study Left-Right Models (LR) is to 
explain the lightness of neutrinos masses. On the literature there are two different Left-Right 
models. They differ in their $SU(2)_{R}$ breaking fields: one uses $SU(2)_{R}$ triplets (LRT) and the
other $SU(2)_{R}$ doublets (LRD). 

However, on the technical side, 
the LR has a problem similar to
that in the SM: the masses of the fundamental Higgs scalars diverge
quadratically. Terefore, we can impose supersymmetry in order to stabilize the scalar
masses and cure this hierarchy problem, as we have done in MSSM.  

The supersymmetric versions of these models, are known as (SUSYLR), have the additional
appealing characteristics of having automatic R-parity conservation. Of course, 
there are two differents kind of model, the first one is the SUSYLRT \cite{susylr}, which is 
the supesymmetric version of LRT model, and the SUSYLRD \cite{doublet}.

Some other possibility of physics beyond the SM, at energies of a few TeVs, is that the gauge
symmetry may be $SU(3)_{c}\otimes SU(3)_{L}\otimes U(1)_{N}$ 
(3-3-1 for shortness).  There are two main versions of the 3-3-1 models as far as lepton
sector is concern. In the minimal version, the charge conjugation
of the right-handed charged lepton for each generation is combined
with the usual $SU(2)_L$ doublet left-handed leptons components to
form an $SU(3)$ triplet $(\nu, l, l^c)_L$.  No extra leptons are
needed in this model, and we shall call such model as minimal 3-3-1 model. We
want to remind that in this model there is no right-handed (RH)
neutrino. There exists  another interesting possibility, where we add a
left-handed
anti-neutrino to each usual $SU(2)_L$ doublet to form an $SU(3)$ triplet
$(\nu, l , \nu^{c})_L$, and this model
is called the 3-3-1 model with RH neutrinos.

The 3-3-1 models \cite{singer,ppf,331rh} provide possible solutions to some puzzles
of the standard model (SM) such as the generation number problem,
the electric charge quantization \cite{dongl2}. Since one
generation of quarks is treated differently from the others this
may  lead to a natural explanation for  the large mass of the top
quark~\cite{longvan}. There is also a good candidate
for self-interacting dark matter (SIDM) since there are two Higgs
bosons, one scalar and one pseudoscalar, which have the properties
of candidates for dark matter like stability, neutrality and that
it must not overpopulate the universe~\cite{longlan}, etc.

As happens with the SUYLR models, again, we have two kind of supersymmetrics model. The 
first one is the Minimal Supersymmetric 3-3-1 model (MSUSY331), the supersymmetric version of the  
minimal 3-3-1 model. The second model is the susy331rh, wchich contains right handed neutrinos. 

The models 3-3-1 can be embedded in a 
model with 3-4-1, its mean $SU(3)_{c}\otimes SU(4)_{L}\otimes U(1)_{N}$ gauge symmetry. The $SU(3)_{L}$ symmetry is
possibly the largest symmetry  involving the known leptons (and $SU(4)_{L}$ if
right-handed neutrinos do really exist). This make 3-4-1 model
interesting by its own. Someyears ago was presented the supersymmetric 
version of these models listed above \cite{331susy,331susy2,susy341}.

By another hand, the Linear colliders would be most 
versatile tools in experimental high energy physics. A large 
international effort is currently under way to study the technical feasibility and physics 
possibilities of linear $e^{+}e^{-}$ colliders in the TeV range. A number of designs have already 
been proposed (NLC, JLC, TESLA, CLIC, VLEPP, ...) and several workshops have recently been devoted 
to this subject. They can provide not only $e^{+}e^{-}$ collisions and high luminosities, but also very 
energetic beams of real photons. One could thus exploit $\gamma \gamma$, $e^{-} \gamma$ and  even 
$e^{-}e^{-}$ collisions for physics studies. Thus it has been proposed to build a new electron-positron 
collider, the International Linear Collider (ILC) \cite{ilc1,ilc2}.

The last exciting prospects have prompted a growing number of theoretical studies devoted to the 
investigation of the physics potential of such $e^{-}e^{-}$ accelerator experiments.  Of course, in the 
realm of the Standard Model this option is not particularly interesting because mainly M\o ller 
scattering, the total cross section to this process is $\sigma \approx 10^{-3}nb$ at $\sqrt{s}=500$ GeV 
\cite{assi3}, and bremsstrahlung events are to be observed. 

However, it is just for that reason that 
$e^{-}e^{-}$ collisions can provide crucial information on exotic processes, in particular on processes 
involving lepton and/or fermion number violation. Therefore, new perspectives emerge in detecting new 
physics beyond the Standard Model in processes having non-zero initial electric charge (and non-zero 
lepton number) like in electron-electron $e^{-}e^{-}$ process.

The goal of this article is to review the mecanism of production of double charged charginos and neutralinos in 
electron-electron process on the supersymmetric models listed above.

\section{Charginos Production}

The Left-Right models may have doubly charged Scalars \cite{cmmc}. It means that, when we construct their 
supersymmetric version, we get double charged charginos. There are another kinds of model, where similar 
situation occur.  Models with  $SU(3)$ (or $SU(4)$) electroweak symmetry may have doubly charged 
vector bosons. This means that in some supersymmetric extensions of these kind 
of models we will have double charged charginos \cite{mcr,susy341}. 

By another way, there are not so many studies about this kind of particle. Due this fact there are 
not experimental studies to detect this kind of particle. Due this fact, here I want to 
summarize the main results in the literature concerning the production of double charged charginos.

In order to start this study, it is useful to review the particle content of which model we have discussed above. 
Instead to present all the particles of each model, on the table 1, we list the 
\begin{table}
\begin{center}
\begin{tabular} {|c|c|}\hline
model    & charginos and neutralinos \\ \hline 
MSSM \cite{mssm} & $\tilde{\chi}^{\pm}(2) \,\ \tilde{\chi}^{0}(4)$ \\ \hline
SUSYLRT \cite{susylr} & $\tilde{\chi}^{\pm \pm}(1) \,\ \tilde{\chi}^{\pm}(5) \,\ 
\tilde{\chi}^{0}(9)$ \\ \hline
SUSYLRD \cite{doublet} & $\tilde{\chi}^{\pm}(6) \,\ \tilde{\chi}^{0}(11)$ \\ \hline
MSUSY331 \cite{331susy} &   $\tilde{\chi}^{\pm \pm}(5) \,\ \tilde{\chi}^{\pm}(8) \,\  
\tilde{ \chi}^{0}(13)$ \\ \hline
SUSY331RN \cite{331susy2}& $\tilde{ \chi}^{\pm}(6) \,\ \tilde{ \chi}^{0}(15)$  \\ \hline
SUSY341 \cite{susy341}& $\tilde{\chi}^{\pm \pm}(5) \,\ \tilde{\chi}^{\pm}(16) \,\  \tilde{\chi}^{0}(25)$ \\ \hline
NMSSM \cite{dress} & $\tilde{\chi}^{\pm}(2) \,\ \tilde{\chi}^{0}(5)$ \\ \hline 
\end{tabular}
\end{center}
\label{t1}
\begin{center}
Table1: Spectrum of Charginos and Neutralinos in several SUSY models
\end{center}
\end{table}
particle content of the chargino's and neutralino's sector at some supersymmetric models. In parenthesis we show the number of 
states that they appear in each models. Therefore we can distinguish the differents models with base in the numbers 
of particles.

As we mentioned above, because of low level of SM backgrounds, the total cross section 
$\sigma \approx 10^{-3}nb$ at $\sqrt{s}=500GeV$ \cite{assi3}, $e^-e^-$ 
collisions are a good reaction for discovering and investigating new physics 
at linear colliders. With this process is possible to study reactions that 
violate both lepton and/or fermion number, and this kind of reaction are expected in 
supersymmetric models, as we will briefelly present next.

Before, we present our review, it is useful to remeber that \cite{dress} ``Sleptons are likely 
to be among the ligther sparticles whose early discovery is anticipated. As already shown, a 
knowledge of the mass parameters $m_{\tilde{l}_{L}}$, $m_{\tilde{l}_{R}}$ and 
$m_{\tilde{\nu}_{L}}$ will be of great use in studying signals of charginos and neutralinos". Selectron 
pair-production takes place in $e^{-}e^{-}$ collisions via the exchange of the neutralinos $\tilde{\chi}^{0}$ 
in a $t$-channel contribution was studied at \cite{cuy,cuy1}. This production depends very crucially
on the properties of the exchanged neutralinos, {\em i.e.} their masses and their couplings to electrons,
because strong interferences can take place between the different channels
and dramatically influence the production cross section. 

\begin{figure}
\begin{center}
\boldmath
\unitlength1mm
\SetScale{2.837}
\mbox{\epsfig{file=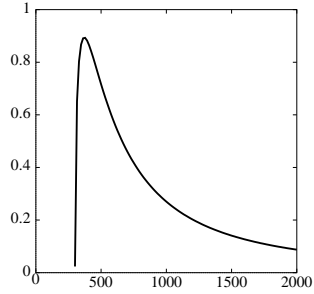,height=0.3\textwidth,angle=0}}
\caption{The total cross section to the selectron production ($\sigma$ [pb]) in $e^-e^-$ collisions as 
  function of the energy ($\sqrt{s_{ee}}$ [GeV]).}
\label{f5}
\end{center}
\end{figure}

\begin{figure}
\begin{center}
\boldmath
\unitlength1mm
\SetScale{2.837}
\mbox{\epsfig{file=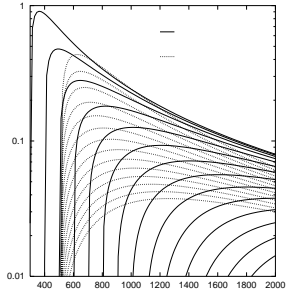,height=0.3\textwidth,angle=0}}
\caption{Energy dependence of the unpolarized production cross sections of
$e^-e^-\to \tilde{e}^{-}\tilde{e}^{-}$ (full curves)
and
$e^-e^-\to \tilde{\chi}_{1}^{-}\tilde{\chi}_{1}^{-}$ (dotted curves)
for $m_{\tilde{e}}=m_{\tilde{\nu}}=$150, 200, \ldots\ 800 GeV,
assuming
$\tan \beta=10$, $\mu=-300$ GeV and $M_{2}=300$ GeV.
For this choice of parameters,
$m_{\tilde{\chi}^{-}_{1}}=255$ GeV.}
\label{f05}
\end{center}
\end{figure}

It is important to note that, this reaction violates fermion number conservation,
which comes as no surprise since the neutralinos are Majorana fermions. On these references cited above, 
the authors studied the cross section to produce the sleptons. Some of their results are depicted in Fig.(\ref{f5},\ref{f05}). 
From this figure, we can notice that the cross section to the selectron production are of the same magnitude 
as the cross section to the M\o ller production. 

However, after impose rapidity, energy and acoplanarity cuts \cite{cuy1} 
the background from M\o ller scattering is entirely eliminated. The supersymmetric signal,
on the other hand, is not significantly reduced by these mild cuts, which roughly simulate a typical detector acceptance. 
Therefore, we can conclude that the $e^{-}e^{-}$ machine is ideal for discovering and studying selectrons.

%\begin{figure*}
%\begin{center}
%\vglue -0.009cm
%\mbox{\epsfig{file=eesingchargmssm.eps,width=2.2in}}
%figure F. Cuypers, G. J. van Oldenborgh, R. Rückl, NPB409, 128 (1993) 
%\caption{Lower Diagram Contributing to $e^{-}e^{-}\to \chi^{-}_{1}\chi^{-}_{1}$ in the MSSM.}
%\label{mssm01} 
%\end{center}     
%\end{figure*}

\begin{figure*}
\begin{center}
\vglue -0.009cm
\mbox{\epsfig{file=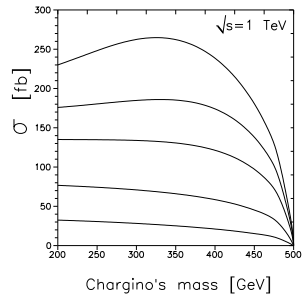,width=0.3\textwidth,angle=270}}
%figure from J. Gluza, T. Wohrmann, PLB408, 229 (1997)
\caption{Cross section for the process
$e^-e^- \rightarrow \tilde \chi^-_1 \tilde \chi^-_1$
as a function of the chargino mass and sneutrino masses
(downwards) $m_{\tilde \nu}=100,200,300,500,800$ GeV and
$\sqrt{s}=1$ TeV.}
\label{mssm2}       
\end{center}
\end{figure*}

In the realm of the MSSM, chargino pairs can be produced in $e^{-}e^{-}$ collisions
by the u- and t-channel exchange of a sneutrinos, as shown at \cite{cuy,glusa}. 
This production depends very crucially on the properties of the exchanged sneutrinos, 
{\em i.e.} their masses and their couplings to electrons-charginos. On this works, the 
authors calculate the total cross section of the reaction $e^{-}e^{-}\to \chi^{-}_{1}\chi^{-}_{1}$ 
in the MSSM. The main results are show on Fig.(\ref{f05},\ref{mssm2}) for unpolarized beams. 
From Fig.(\ref{mssm2}), that the cross section of the production of the charginos are bigger than 
cross section to the M\o ller production for several values of charginos masses.

\begin{figure*}
\begin{center}
\vglue -0.009cm
\mbox{\epsfig{file=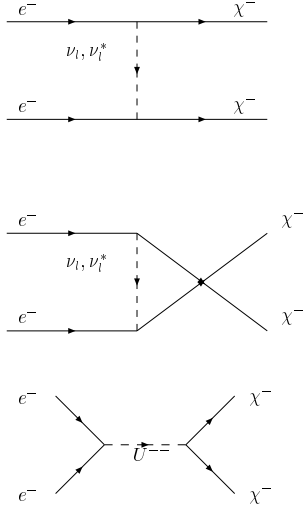,width=0.3\textwidth,angle=0}}
\caption{Lower Diagram Contributing to $e^{-}e^{-}\to \chi^{-}_{1}\chi^{-}_{1}$ in the SUSY331 and SUSY341.}
\label{figsusy331-1}       
\end{center}
\end{figure*}

\begin{figure*}
\begin{center}
\vglue -0.009cm
\mbox{\epsfig{file=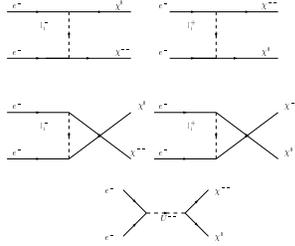,width=0.3\textwidth,angle=0}}
\caption{Lower Diagram Contributing to $e^{-}e^{-}\to \chi^{--}_{1}\chi^{0}_{1}$ in the SUSY331 and SUSY341.}
\label{figsusy331-2}       
\end{center}
\end{figure*}

While in the case of the supersymmetric 331 and 341 model, the Feynmann diagrams 
contributing to $e^-e^- \to \tilde{ \chi}^{-} \tilde{ \chi}^{-}$ is 
shown in Fig(\ref{figsusy331-1}). The Feynmann diagram that contribute 
to the $e^-e^- \to \tilde{ \chi}^{--} \tilde{ \chi}^{0}$ is show 
in the Fig.(\ref{figsusy331-2}). We must stress that in the MSSM 
the chargino pairs can be produced in $e^-e^-$ collisions by the u- and t- 
channel exchange of a sneutrino. In both model, susy331 and susy341, we have beyond this possibility, 
the s- channel contributing with the exchange of a bilepton $U^{--}$, because 
of this new contribution we have on peak at $\sqrt{s} \simeq M_U$, where $M_U$ 
is the bilepton mass is expected. The total cross section outside the $U$'s resonance has 
the same order of magnitude than the cross section in the MSSM.

\begin{figure}[ht]
\begin{center}
\vglue -0.009cm
\mbox{\epsfig{file=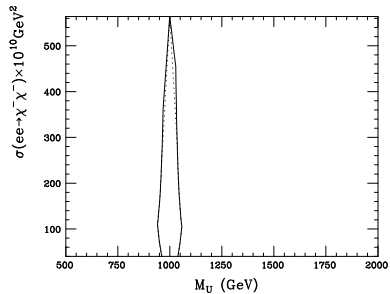,width=0.3\textwidth,angle=90}}       
\end{center}
\caption{Total Cross Section $e^-e^- \to \tilde{ \chi}^{-} \tilde{ \chi}^{-}$ 
at $\sqrt{s}=1.0$TeV in susy331 and susy341 models.}
\label{plot2}
\end{figure}

The cross section of these process was calculated on \cite{mcr}, and the total cross section is show in the 
Fig.(\ref{plot2}). The results is that outside the $U$ resonance, the total cross section is of 
order of pb, like in the MSSM, and near the $U$ resonance we have very nice peak. Due to this 
fact we expect that there will be an enhancement in the cross 
section of production of these particles in $e^-e^-$ collisors, such as the ILC~\cite{mcr}.

\begin{figure*}
\begin{center}
\vglue -0.009cm
\mbox{\epsfig{file=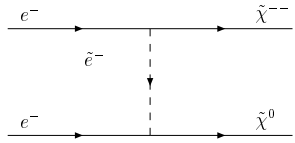,width=0.3\textwidth,angle=0}} 
\caption{Lower Diagram Contributing to $e^{-}e^{-}\to \chi^{--}_{1}\chi^{0}_{1}$ in the SUSYLRT.}
\label{figsusylr1}      
\end{center}
\end{figure*}

The production of double charged chargino in $e^+e^-$ collision occurs through the diagrams
presented in Fig.(\ref{figsusy331-2}) on the models susy331 and susy341. While on Fig.(\ref{figsusylr1}) we present 
the Feynmann diagram to this process on SUSYLRT. Comparing Figs.(\ref{figsusy331-2},\ref{figsusylr1}) we notice that 
in the models, susy331 and susy341, have one contribution on $s$-channel that don't appear on the SUSYLRT.

\begin{figure*}
\begin{center}
\vglue -0.009cm
\mbox{\epsfig{file=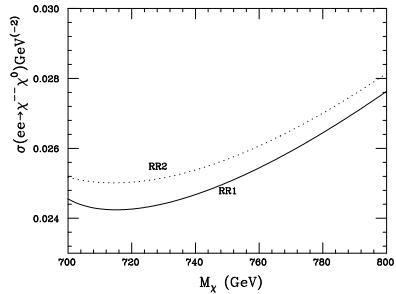,width=0.3\textwidth,angle=90}}
\caption{Total cross section to $e^{-}e^{-}\to \chi^{--}_{1}\chi^{0}_{1}$ in the SUSY331 and SUSY341 as 
fuction of the double charged mass.}
\label{figsusy331-3}       
\end{center}
\end{figure*}

\begin{figure*}
\begin{center}
\vglue -0.009cm
\mbox{\epsfig{file=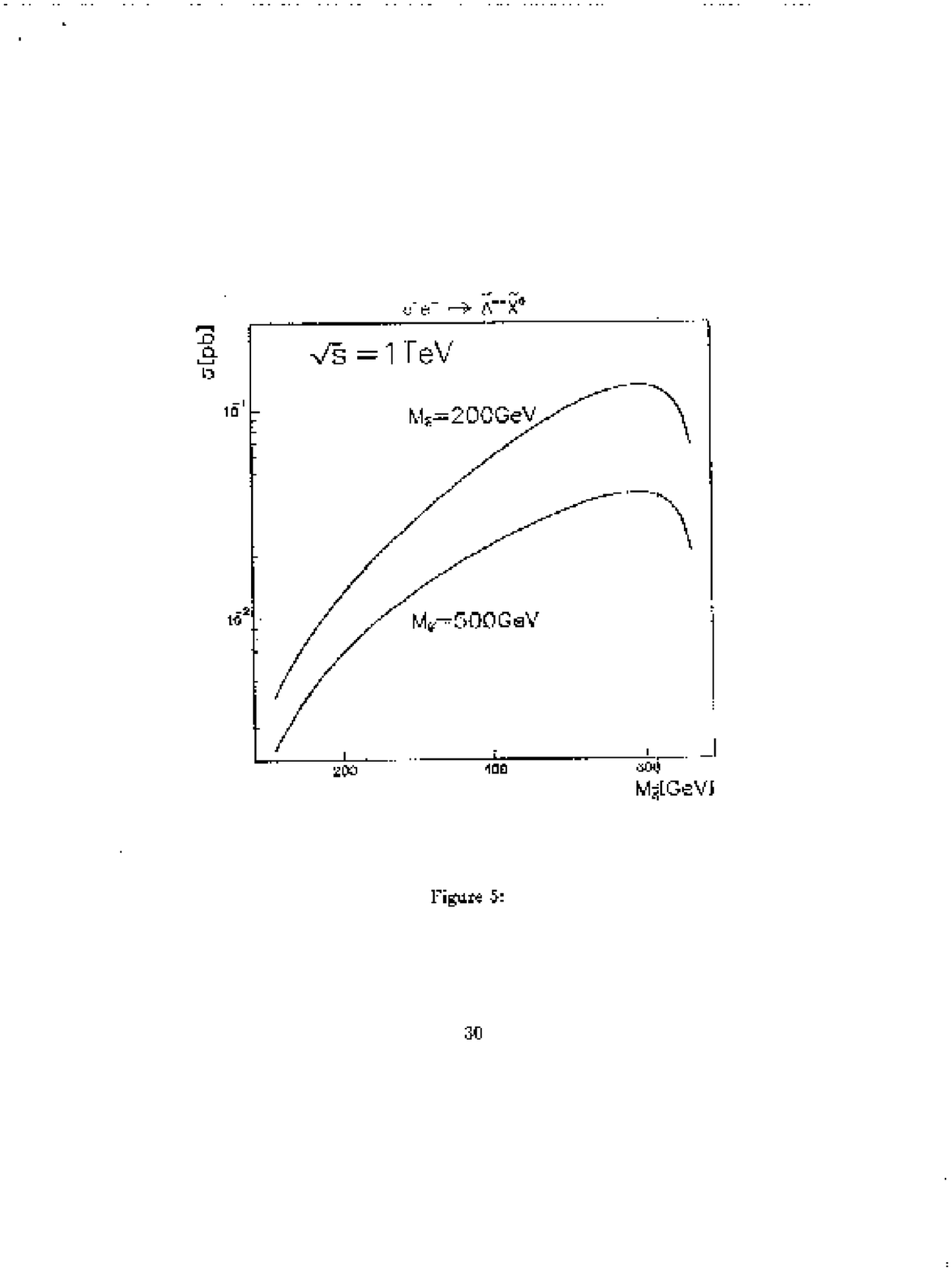,width=0.6\textwidth,angle=0}} 
\caption{Total cross section to $e^{-}e^{-}\to \chi^{--}_{1}\chi^{0}_{1}$ in the SUSYLRT  as 
fuction of the double charged mass.}
\label{figsusylr2}      
\end{center}
\end{figure*}

The total cross section to this process was calculated 
on \cite{huitu}, and we show on Fig(\ref{figsusylr2}) the cross section as function of the mass of the double 
charged chargino. The results on both, susy331 and susy341 models, are presented at Fig.(\ref{figsusy331-3}). We notice 
that allways the cross section in susy331 and susy341 model is greater than the ones get at SUSYLRT.

We have considerate the double chargino mass in the range $700 \leq 
M_{\tilde{\chi}^{++}} \leq 800$ GeV, and we could get cross section of the 
order of pb outside the $U$ resonance, while in the resonance we 
have an enhancement in the cross section. We believe that these new states can be 
discovered, if they really exist, in linear colliders

\section{Conclusions}
\label{sec:con}

We believe that the charginos and neutralinos production can be very well studied 
in the international linear colliders (ILC). Due the fact that the differents models presented here 
have differents predictions, on the mechanism production, they can distinguish at ILC. Another exciting 
search, can be done in discover the double charged charginos, due the fact that there are very few models that 
predict these kind of particle, and if they really exist the ILC can detect them.

\section{acknowledgements}

This work was supported by Conselho Nacional de Ci\^encia e Tecnologia (CNPq) 
under the processes 309564/2006-9.


\begin{thebibliography}{99}
\bibitem {mssm}H. E. Haber and G. L. Kane, {\sl Phys. Rep.}{\bf 117}, 75 (1985).
\bibitem{INO82a} K. Inoue, A. Komatsu and S. Takeshita,
 Prog. Theor. Phys. {\bf 68} (1982) 927.
\bibitem{INO82b}K. Inoue, A. Komatsu and S. Takeshita,
 Prog. Theor. Phys. {\bf 70} (1983) 330.
\bibitem{haber2} H. E. Haber, Eur. Phys. J. C
{\bf15}, 817 (2000).
\bibitem {susylr}R. M. Francis, M. Frank and C. S. Kalman, {\sl Phys. Rev.}
{\bf D43}, 2369 (1991); C.S.Aulakh,A.Melfo and G.Senjanovic,
{\sl Phys.Rev.}{\bf D57},4174 (1998); G. Barenboim and N. Rius,
{\sl Phys. Rev.}{\bf D58}, 065010, (1998); N. Setzer and S. Spinner,
{\sl Phys. Rev.} {\bf D71}, 115010 (2005).
\bibitem {doublet}K. S. Babu.B. Dutta and R.N. Mohapatra, {\sl Phys.Rev.}
{\bf D65},016005, (2002).
\bibitem{singer}M. Singer, J. W. F. Valle, and J. Schechter, Phys.
Rev. D {\bf22}, (1980) 738
\bibitem{ppf} F. Pisano and V. Pleitez, Phys. Rev.  D {\bf46}, (1992) 410;
P. H. Frampton, Phys. Rev. Lett. {\bf69}, (1992) 2889; R. Foot
{\it et al,} Phys. Rev. D {\bf47}, (1993) 4158.
\bibitem{331rh} R. Foot, H. N. Long, and Tuan A.
Tran,
 Phys. Rev. D {\bf50}, (1994) R34; J. C. Montero, F. Pisano, and
 V. Pleitez, Phys. Rev. D {\bf 47}, (1993)
 2918; H. N. Long, Phys. Rev. D  {\bf 53}, (1996) 437;
Phys. Rev. D  {\bf 54}, (1996) 4691.
\bibitem{dongl2}  F. Pisano, Mod. Phys. Lett
A {\bf 11} 2639 (1996); A. Doff and F. Pisano, Mod. Phys. Lett A
{\bf 14} 1133 (1999); C. A. de S. Pires and O. P. Ravinez, Phys.
Rev. D {\bf 58} 035008 (1998); C. A. de S. Pires, Phys. Rev. D
{\bf 60} 075013 (1999);  P. V. Dong and H. N. Long, [arXiv:
hep-ph/0507155].
\bibitem{longvan} H. N. Long and V. T. Van, J. Phys. {\bf G25}, 2319 (1999).
\bibitem{longlan} D. Fregolente and M. D. Tonasse, Phys. Lett. B
{\bf 555}, 7 (2003);
 H. N. Long and N. Q. Lan, Europhys. Lett. {\bf 64}, 571
(2003).
\bibitem{331susy}  H. N. Long and P. B. Pal, {\sl Mod. Phys. Lett.} {\bf A13}, 2355 
(1998);  T. V. Duong and E. Ma, {\sl Phys. Lett.} {\bf B316}, 307 (1993);
J. C. Montero, V. Pleitez and M. C. Rodriguez,
{\sl Phys. Rev.} {\bf D65}, 035006 (2002).
\bibitem{331susy2} J. C. Montero, V. Pleitez and M. C. Rodriguez, 
{\sl Phys. Rev.} {\bf D70}, 075004 (2004).
\bibitem{susy341}M. C. Rodrgiuez, hep-ph/0701088.
\bibitem{ilc1}M. Battaglia,  T. Barklow, M. Peskin, 
Y. Okada, S. Yamashita and P. Zerwa, 
In the Proceedings of 2005 International Linear Collider Workshop (LCWS 2005), Stanford, California, 18-22 Mar 2005.
In the Proceedings of 2005 International Linear Collider Workshop (LCWS 2005), Stanford, California, 18-22 Mar 2005, 
hep-ex/0603010.
\bibitem{ilc2}M. Battaglia, arXiv:0705.3997 [hep-ex].
\bibitem{assi3} J. C. Montero, V. Pleitez and M. C. Rodriguez, 
{\sl Int. J. Mod. Phys. A16}, 1147, (2001).
\bibitem{cmmc}C.M. Maekawa and M. C. Rodriguez, JHEP {\bf 04}(2006),031.
\bibitem{mcr} M. Capdequi-Peyranere and M. C. Rodriguez, {\sl Phys. Rev.} 
{\bf D65}, 035006 (2002).
\bibitem {dress}M. Dress, R. M. Godbole and P. Royr, 
{\it Theory and Phenomenology of Sparticles}
1st edition, World Scientific Publishing Co. Pte. Ltd., Singapore, (2004).
\bibitem{cuy} F. Cuypers, G. J. van Oldenborgh, R. Rückl, {\sl Nucl. Phys.}{\bf B409}, 128 (1993).
\bibitem{cuy1} F. Cuypers, Int.J.Mod.Phys.A11:1585-1590,1996; F. Cuypers, hep-ph/9603243.
\bibitem{glusa} J. Gluza, T. Wohrmann, {\sl Phys. Lett.} {\bf B408}, 229 (1997).
\bibitem{huitu}K.Huitu, J.Maalampi, M. Raidal, {\sl Nucl. Phys.}{\bf B420}, 449 (1994).
\end{thebibliography}
\end{document}